# MerKurio: sequence extraction and annotation based on matching k-mers

Lukas Schönmann, Heinz Himmelbauer, Juliane C. Dohm*

Institute of Computational Biology, Department of Biotechnology and Food Science, BOKU University, Muthgasse 18, 1190 Vienna, Austria

lukas.schoenmann@students.boku.ac.at

heinz.himmelbauer@boku.ac.at

*corresponding author: Juliane Dohm, dohm@boku.ac.at, ORCID 0000-0001-5970-4457

## Keywords

Bioinformatics, *k*-mers, Sequence analysis, Software, Rust programming language

## Abstract

*K*-mers, short subsequences of length *k*, play an important role in computational analyses using biological sequence data. A critical downstream processing step involves getting back to the source sequences of selected *k*-mers for further analysis or validation. Here, we present MerKurio, a high-performance command line tool written in Rust with two main functionalities: 1) extracting sequence records from FASTA/FASTQ files using *k*-mers and 2) annotating or filtering aligned sequences in SAM/BAM files with *k*-mer tags. The tool automatically selects the pattern matching algorithm via empirically derived rules based on query characteristics and supports paired-end reads, compressed input files, and reverse complement searching. Benchmark comparisons demonstrate that MerKurio outperforms existing tools in terms of speed while providing additional utility including detailed matching statistics and comprehensive file format support. MerKurio is a fast and user-friendly tool for sequence record extraction and tagging of aligned sequences based on *k*-mers. MerKurio is freely available at https://github.com/lschoenm/MerKurio.

## Introduction

In recent years, *k*-mer-based methods utilizing substrings of length *k* derived from biological sequences (DNA, protein) have become increasingly important in the field of bioinformatics (Marchet et al., 2021; Moeckel et al., 2024). Initial steps of k-mer analysis, such as counting and frequency estimation, have been well-established through efficient tools (Moeckel et al., 2024) that can process large genomic datasets using optimized memory structures including Bloom filters and minimizers. For similarity estimation, the MinHash technique implemented in tools like Mash enables rapid genome comparisons, and specialized indexing data structures support efficient querying of massive *k*-mer collections (Marchet et al., 2021).

Based on such computational methods, *k*-mer-based approaches have found diverse applications, for example genome size estimation (Fu et al., 2024; Liao et al., 2024), genome assembly (Eghdami et al., 2025; Luo et al., 2012), taxonomic classification in metagenomics (Ye et al., 2019), phylogenetic analysis based on *k*-mer sketches (Ondov et al., 2016; Sandell et al., 2022), as well as the analysis of genomic structural variation (Abo et al., 2015). In genome re-sequencing projects, reference-free *k*-mer-based genotype representations have enhanced the identification of genomic regions linked to specific traits as compared to traditional association studies (Karikari et al., 2023).

All these applications gather *k*-mer-based insights after *k*-mer extraction. However, tracing *k*-mers of interest back to their sequences in the original data is a desired task in many analysis workflows, and this reverse lookup is still lacking highly optimized solutions. For instance, *k*-mer-based genome-wide association studies (GWAS) may identify thousands of trait-associated *k*-mers that require extraction of the corresponding sequencing reads for downstream analysis (Lemay et al., 2023). There is a need for specialized tools designed to process large *k*-mer datasets that are both user-friendly and accessible to broader research communities (Karikari et al., 2023; Lemay et al., 2023).

Here, we present MerKurio, a usability-oriented, high-performance tool designed to link *k*-mers of interest back to their sequences of origin. MerKurio implements the Backward Nondeterministic DAWG Matching tuned with *q*-grams (BNDMq) algorithm (Ďurian et al., 2009) and the Aho-Corasick algorithm (Aho & Corasick, 1975) for substring search, enabling efficient extraction of records from genomic, transcriptomic, and protein sequence files. Comprehensive output on matching statistics of the query *k*-mers is an (optional) additional feature. Furthermore, MerKurio provides a function to tag aligned sequences with the *k*-mers they contain. Various input file formats and different *k*-mer lengths are supported.

# Materials and Methods

MerKurio was written in Rust version 1.90.0-nightly. Automated building and publishing of binaries was implemented with dist v0.28.6 (https://github.com/axodotdev/cargo-dist). Online documentation was generated using mdbook v0.4.40 (https://github.com/rust-lang/mdBook).

MerKurio relies on efficient pattern matching to identify *k*-mer positions in sequences. We implemented BNDMq as described by Ďurian et al. (Ďurian et al., 2009), the Aho-Corasick algorithm (Aho & Corasick, 1975) via the aho_corasick Rust crate (https://github.com/BurntSushi/aho-corasick), and a rolling-hash matcher using a 64-bit polynomial fingerprint. Automatic algorithm selection is based on pattern length, the number of patterns, and whether searching stops at the first match or finds all occurences. We systematically compared the three implementations using simulated DNA patterns and 150 bases long sequences. The algorithm tuning covered pattern lengths from 4 to 128 and pattern counts up to one million. Our tests have shown that BNDMq was generally fastest for small sets of patterns that fit within a processor word (64 bits), Aho-Corasick was best for intermediate pattern counts and mixed-length queries, and rolling hash matching perfromed best for larger sets of equal length patterns. Decompressing and parsing FASTA/FASTQ data was implemented using the Rust library paraseq (https://github.com/noamteyssier/paraseq), and pre-processing was performed with needletail (https://github.com/onecodex/needletail).

Test data of paired-end *Staphylococcus aureus* genomic sequencing reads were downloaded from https://gage.cbcb.umd.edu/data/Staphylococcus_aureus/Data.original/ (Salzberg et al., 2011) (frag_1.fastq.gz, frag_2.fastq.gz; SRA identifier SRR022868), and sugar beet test data were taken from SRA Bioproject PRJNA815240 (Sandell et al., 2022). Test *k*-mers were extracted from the sequencing data using awk. Benchmarks on the *S. aureus* data were run and recorded using hyperfine v1.18.0 (https://github.com/sharkdp/hyperfine), with 20 warm-up runs followed by 100 benchmarking runs, each pinned to the same CPU given the highest CPU priority. Benchmarking using *S. aureus* data was performed on a standard Linux computer with an AMD Ryzen 5 5600X CPU (6 cores, 3.7 GHz), 16 GB DDR4 RAM, and an NVMe SSD (Crucial P3 2 TB, model CT2000P3SSD8) running Ubuntu 20.04.4 LTS. The tests using 1.3 TB of sugar beet data were repeated several times on a high-performance computing node running under AlmaLinux 8.10 with 64 cores (3.8 GHz) and 768 GB memory, the run time using 32 cores (multithreaded with GNU parallel (Tange, 2018)) was recorded using the Linux command "time".

Pre-compiled binaries of MerKurio are provided for multiple architectures at https://github.com/lschoenm/MerKurio/releases and can be executed from the command line in Unix-like shell terminals on all common operating systems (Linux, MacOS, Windows) with "merkurio extract -i in.fa -f kmers.txt -o out.fa". Comprehensive documentation is available including example data.

# Results

MerKurio was developed as command line tool and provides two main functionalities through dedicated subcommands. The "extract" subcommand identifies and extracts sequence records from FASTA or FASTQ files based on query *k*-mers, with full support for paired-end reads (match in one read extracts both mates). The "tag" subcommand annotates sequences in SAM or BAM alignment formats according to contained *k*-mers using user-defined labels (default “km” followed by the comma-separated *k*-mers) in compliance with the SAM Optional Fields Specification (The SAM/BAM Format Specification Working Group, 2024).

Both subcommands offer detailed match statistics (positions, counts, summary statistics, metadata), support for compressed input files (gzip, bzip2, xz), the option to extract records without matches, reverse complement searching, support of all possible letters in DNA and protein sequences (incl. "N", interpreted as a regular search character), support of different *k*-mer lengths in the same input, automatic recognition of input/output formats, and statistics export in plain text or JSON format. Query *k*-mers can be provided via command line arguments or file input. Native parallelization enables the utilization of multiple CPU cores for accelerated processing.

MerKurio was implemented in Rust due to its memory safety and performance.

## Comparisons with existing tools

We compared MerKurio to seven tools that may be used for the same purpose including grep (GNU Project, 2020), seqtool (https://github.com/markschl/seqtool), seqkit (Shen et al., 2024), cookiecutter (https://github.com/ad3002/Cookiecutter), fetch_reads (https://github.com/voichek/fetch_reads_with_kmers), katcher (Lemay et al., 2023), and back_to_sequences (Baire et al., 2024). All these programs can in principle extract sequences based on matching *k*-mers but differ considerably in terms of applicable input/output formats, search process and performance, and extent of output information (Table 1). We used back_to_sequences version 0.7.0 instead of version 0.8.4 because the latest release does not support FASTQ output.

The text searching tool grep, written in C and standard on Unix-like systems, performs best when ignoring regular expression patterns using the -F flag (or running it as "fgrep"). While fast, its generality implies that it does not recognize biological file formats. Seqtool and seqkit are multi-purpose bioinformatics tools specialized on sequence data but do not support SAM/BAM files nor the extraction of paired reads. Cookiecutter filters reads based on a *k*-mer library and can handle paired-end reads in parallel but is limited to processing uncompressed FASTQ files. Fetch_reads is optimized for fast extraction of paired reads but imposes constraints on the input: *k*-mers must be provided in FASTA format with $k < 32$. Katcher is restricted to data in BAM-format and allows for tagging and extracting aligned reads based on *k*-mers but only supports a fixed *k*-mer length determined at compile time. Both fetch_reads and katcher always search for reverse complements, and neither tool can handle "N" characters nor others apart from A, C, G, T. Back_to_sequences was developed specifically to find the origin of *k*-mers, and match positions within sequences are reported. However, the program does not accept multiline FASTA nor paired-end FASTQ as input, and its output is not well suited for automated parsing.

MerKurio, in contrast, works on both single-end and paired-end data in compressed or uncompressed FASTA or FASTQ format at flexible *k*-mer lengths and provides comprehensive information (as option) along with the extracted sequences. Additionally, MerKurio can perform annotation of aligned sequences in SAM and BAM format.

### Performance benchmarks

To assess the performance of pattern matching and record processing, we conducted benchmarks on raw sequencing data and compared the “extract” subcommand of MerKurio to six out of seven tools mentioned above that can handle raw data (omitting katcher that uses alignment files).

A dataset of paired Illumina short reads from *Staphylococcus aureus* was selected from the Genome Assembly Gold-Standard Evaluations (GAGE) study collection (Salzberg et al., 2011). Its relatively small size (647,052 read pairs, each read 101 basepairs long) allowed for efficient repeated benchmark runs while representing realistic genomic complexity.

One and 100 query *k*-mers were used to search reads either in a single FASTQ file containing the first read of a paired-end data set or in two paired FASTQ files containing read 1 and read 2, respectively. The query *k*-mers were sampled from these sequencing data with *k*=31, a common length for *k*-mer-based methods (Lemay et al., 2023; Voichek & Weigel, 2020) and at the same time the upper limit of the tool fetch_reads.

MerKurio’s average execution times generally outperformed all tested tools across different scenarios (Table 2). When searching for a single *k*-mer in a single input file the execution times of fgrep and seqtools were slightly slower than MerKurio but searching 100 *k*-mers was four times (fgrep) or twelve times (seqtools) slower, respectively. Searches on both strands in paired-end reads including reverse complements using MerKurio were around five times faster than fetch_reads and around elevent times faster than cookiecutter for a single query pattern. In the 100-query paired-end case, however, fetch_reads was slightly faster than MerKurio (0.510 versus 0.541 s). When searching for a million *k*-mers and their reverse complements in paired-end reads, MerKurio was again fastest. The other tools did not support paired-ends in FASTQ format as input files.

In a second benchmark experiment we searched 1.3 terabytes of compressed paired-end sequencing reads from 480 plant genomes (Sandell et al., 2022) for 100 randomly chosen *k*-mers of length 31 and their reverse complements. MerKurio completed the read extraction in 12.9 minutes while also recording the exact positions of each *k*-mer occurrence per sample. The same task took 20.5 minutes using fetch_reads. Repeated tests varied within the range of one minute.

## Conclusions

MerKurio is a fast and user-friendly tool developed in Rust for finding *k*-mers in biological sequences with informative output and comprehensive documentation. The two main functionalities are “extract” for record extraction from FASTA/FASTQ files and “tag” for annotation and filtering of SAM/BAM alignment files. MerKurio's flexibility makes it easy to be integrated into various bioinformatics analysis pipelines using workflow systems or shell scripts. The matching statistics may even serve as features for machine learning models in clustering or classification tasks, like pathogen detection and trait prediction applications. Our results demonstrate that MerKurio effectively fills a gap in current *k*-mer processing tools by offering robust performance and advanced features.

# Acknowledgements

We thank Thomas Holzweber for input regarding the algorithms and tools used for comparison and Christian Gottschall for Linux cluster administration. The large language model GPT-5 by OpenAI was used for polishing of the manuscript draft and assisting code writing (refactoring, bug fixing). The final version of the manuscript was written without AI-writing tools.

# Conflict of interest

The authors declare that they have no competing interests.

# Funding

This work was supported by the Austrian Science Fund (FWF) grant number PAT4292424 "BeetSV". Additional support to L.S. was provided by the Doctoral School AgriGenomics of BOKU University.

# Data availability

MerKurio source code and pre-compiled binaries are available at https://github.com/lschoenm/MerKurio. Installation and usage are documented at https://lschoenm.github.io/MerKurio. A test data set is available. No new data were generated or analyzed in support of this research.

# Tables

**Table 1**. Features of MerKurio and similar programs.

| | MerKurio | back_to_sequences | seqtool | seqkit | katcher | fgrep | fetch_reads | cookiecutter |
|---|---|---|---|---|---|---|---|---|
| Version (year) | 1.1.1 (2026) | 0.7.0 (2025) | 0.4.0-beta.3 (2026) | 2.14.0 (2026) | 0.1 (2024) | 3.11 (2023) | 0_1_beta (2020) | 1.0.0 (2016) |
| Programming language | Rust | Rust | Rust | Go | C | C | C++ | C++ |
| Extracting records from a single FASTQ file | + | + | + | + | - | + | - | + |
| Supporting paired-end reads | + | - | - | - | - | - | + | + |
| Extracting records from FASTA | + | +[a] | + | + | - | +[a] | - | - |
| Detailed matching statistics | + | + | - | - | - | - | - | - |
| Accepting compressed sequence files as input | + | + | + | + | - | - | + | - |
| Searching for reverse complements of *k*-mers | + | + | - | + | + | - | + | - |
| Searching for forward *k*-mers only | + | + | + | + | - | + | - | + |
| Tagging and filtering SAM/BAM files | + | - | - | - | + | - | - | - |
| Built-in parallelization | + | + | + | + | + | - | - | - |

[a]does not support multiline FASTA

**Table 2**. Performance comparison for searching one, 100 or 1,000,000 *k*-mers in a single and paired FASTQ files. Average execution times and standard deviations of 100 benchmark runs (ten runs for 1,000,000 *k*-mers) are reported in seconds with relative performance compared to MerKurio.

| | Single FASTQ | | | | | | Paired FASTQ | | | | | |
|---|---|---|---|---|---|---|---|---|---|---|---|---|
| No. of query *k*-mers | 1 | | 100 | | 1,000,000 | | 2[a] | | 200[a] | | 2,000,000[a] | |
| | Time [s] | Factor | Time [s] | Factor | Time [s] | Factor | Time [s] | Factor | Time [s] | Factor | Time [s] | Factor |
| MerKurio | 0.045 ± 0.004 | 1.00x | 0.189 ± 0.004 | 1.00x | 2.499 ± 0.860 | 1.00x | 0.097 ± 0.002 | 1.00x | 0.541 ± 0.007 | 1.00x | 1.581 ± 0.105 | 1.00x |
| back_to_sequences | 1.072 ± 0.027 | 23.82x | 1.243 ± 0.081 | 6.58x | 3.066 ± 0.034 | 1.23x | - | - | - | - | - | - |
| seqtool | 0.065 ± 0.003 | 1.44x | 2.341 ± 0.070 | 12.39x | > 300[b] | - | - | - | - | - | - | - |
| seqkit | 0.138 ± 0.005 | 3.07x | 4.647 ± 0.075 | 24.59x | > 300[b] | - | - | - | - | - | - | - |
| fgrep | 0.084 ± 0.006 | 1.87x | 0.745 ± 0.029 | 3.94x | 10.356 ± 0.579 | 4.14x | - | - | - | - | - | - |
| fetch_reads | - | - | - | - | - | - | 0.501 ± 0.009 | 5.16x | 0.510 ± 0.007 | 0.94x | 2.664 ± 1.466 | 1.69x |
| cookiecutter | 0.518 ± 0.007 | 11.51x | 0.983 ± 0.039 | 5.20x | 18.203 ± 0.318 | 7.28x | 1.071 ± 0.016 | 11.04x | 2.171 ± 0.021 | 4.01x | 50.325 ± 0.849 | 31.83x |

[a]query *k*-mers and their reverse complements
[b]a single run exceeded the 300 s time limit, and no repeated measurements were performed